\documentclass[a4paper,11pt]{article}
\usepackage{jinstpub}
\usepackage{lineno}
\usepackage{comment}
\title{\boldmath 
A method to measure the quenching factor for recoil energy of oxygen in bismuth germanium oxide scintillators
}

\author{Yuga~Ommura$^a$,}
\author{Hiroshi~Ito$^{a,1}$\note{Corresponding author.},}
\author{Takatomi~Yano$^b$,}
\author{Akihiro~Minamino$^c$,}
\author{and Masaki~Ishitsuka$^a$}

\affiliation[a]{Department of Physics, Faculty of Science and Technology, Tokyo University of Science, Noda, Chiba 278-8510, Japan}
\affiliation[b]{Kamioka Observatory, Institute for Cosmic Ray Research, University of Tokyo, Kamioka, Gifu 506-1205, Japan}
\affiliation[c]{Department of Physics, Yokohama National University, Yokohama, Kanagawa, 240-8501, Japan}

\emailAdd{itoh.hiroshi@rs.tus.ac.jp}

\abstract{
Bismuth germanium oxide ($\rm Bi_4 Ge_3 O_{12}$, BGO) scintillation crystals are widely used as detectors in the fields of particle physics and astrophysics due to their high density, and thus higher efficiency for gamma-ray detection. Owing to their good chemical stability, they can be used in any environment. For rare-event searches, such as dark matter and coherent elastic neutrino-nucleus scattering, BGO crystals are essential to comprehend the response of nuclear recoil. In this study, we have analyzed the events of neutron elastic scattering with oxygen in BGO crystals. Then, we have measured the quenching factor for oxygen recoil energy in the BGO crystal as a function of recoil energy by using a monoenergetic neutron source.
}

\keywords{
}

\begin{document}
\maketitle
\flushbottom

\section{Introduction}
\label{sec:intro}

Bismuth Germanium Oxide ($\rm Bi_4 Ge_3 O_{12}$, BGO) crystal is a scintillation material that is widely used as a detector in particle physics and astrophysics. Some of the advantages are rich light yield and large atomic number to enhance detection efficiency of high energy photons. It is also suitable for calorimetry~\cite{DAMPE:2017fbg, BGO-OD:2019utx}. Recently, rare-event searches, such as dark matter~\cite{JLTP2008, CORON20091393} and coherent elastic neutrino-nucleus scattering~\cite{BIASSONI2012151} have been conducted using BGO crystal. In the events of nuclear recoil with lower energy, it should be considered with light emission quench of the recoil energy.

The yield of scintillation light for deposited energy of recoil nuclei in the inorganic crystal is generally smaller than that for deposited energy of recoil electrons. A ratio of the visible energy $E_{\rm vis}$ to the energy of recoil nuclei $E_{\rm R}$ is called quenching factor ($QF$). It is defined as:
\begin{eqnarray}
QF=E_{\rm vis}/E_{\rm R},
\end{eqnarray}
where $QF$ depends on the material and energy of recoil nuclei. In the scintillating bolometric techniques at 20~mK, the $QF$ was determined to be 12–15\% for several dozen keV in BGO crystals, which is the ratio of the relation between light yield and heat signals for gamma-ray and nuclear recoil events~\cite{CEBRIAN200297, CORON20091393, CORON2004159}. In this study, we have reported a method to measure the $QF$ for oxygen recoil in BGO crystal using monoenergetic neutron irradiation that is capable of expanding the recoil energy up to MeV scale of recoil energy.

The rest of the article is organized as follows. In Sec.~\ref{sec:exp}, we have briefly described the experimental setup and detector calibration. In Sec.~\ref{sec:sim}, we have explained the details of simulation and compared the results with that of the measured spectrum. Analysis of the proposed method and summary of results for $QF$ are shown in Sec.~\ref{sec:ana} and \ref{sec:result}, respectively. Finally, we have presented the conclusion.

\section{Experiment}
\label{sec:exp}

\subsection{Setup}
\begin{figure}[htbp]
    \centering 
    \includegraphics[width=\textwidth]
    {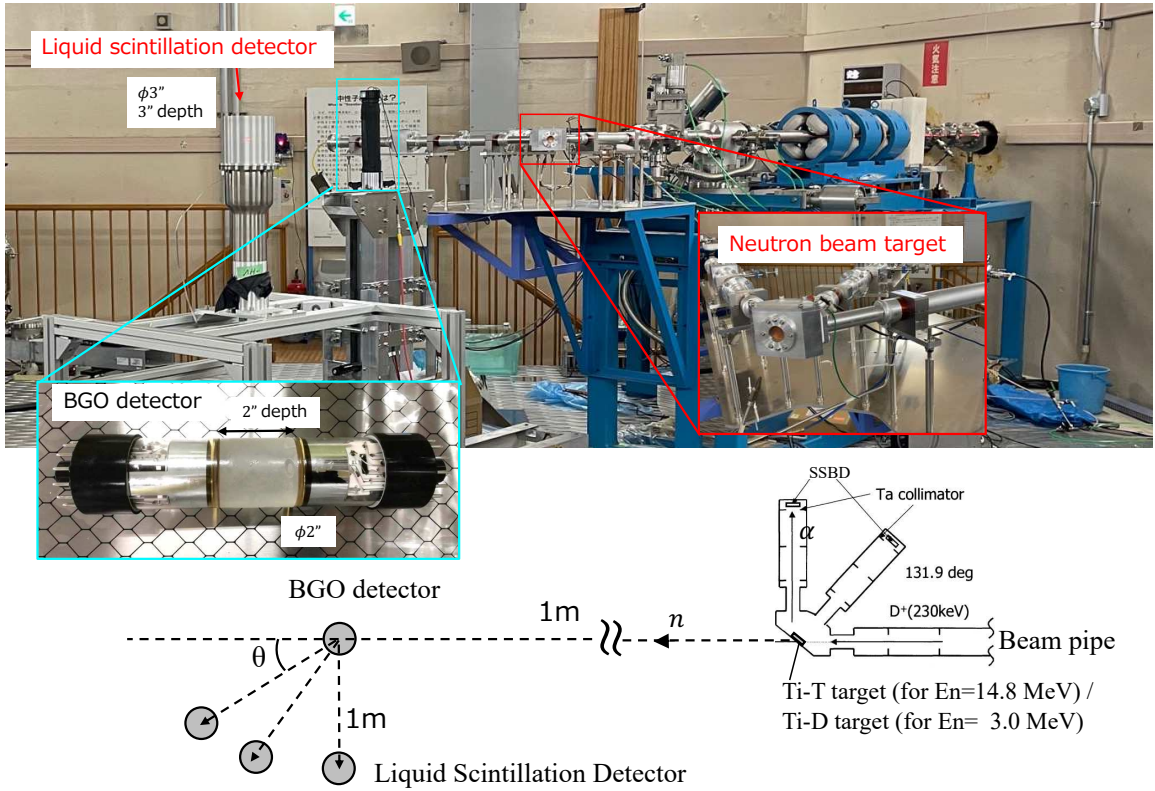}
    \caption{\label{fig:beamtest}
    Graphical view (top) and schematic view (bottom) of setup for neutron irradiation at AIST, Japan.
    The neutron scattering angle $\theta$ is varied up to 90$^{\circ}$.
    }
    \label{figure.setup}
    \end{figure}

To measure the $QF$ of BGO, we have performed an experiment with accelerator-based neutron facility at the National Institute of Advanced Industrial Science and Technology (AIST), Japan~\cite{AIST.2009}. Figure~\ref{fig:beamtest} shows the graphical and schematic view of our experimental setup. The monoenergetic neutrons are produced by deuterium interacting with a titanium-tritium target (for 14.8~MeV) and titanium-deuterium target (for 3 MeV) through the process of T($d,n$)$^4$He and $d$($d,n$)$^3$He, respectively. The neutron fluence is a few hundred $\rm cm^{-2}$. To monitor the 14.8~MeV neutron fluence, the associated alpha particles produced by T($d,n$)$^4$He toward 90.7$^{\circ}$ and 131.9$^{\circ}$ are collimated with tantalum and detected by silicon surface barrier detectors (SSBD). To monitor the 3.0~MeV neutron fluence, a proton recoil is detected by a 0.5~mm thick polyethylene disk mounted SSBD. A BGO detector is placed at the center of the deuterium beamline at a distance of 1~m from the Ti-T (Ti-D) target and liquid scintillation (LS) detector is placed at a distance of 1~m from the BGO detector by changing the scattering angle $\theta$ defined in Fig.~\ref{fig:beamtest}. The BGO detector is set to measure the quenching factor for nuclei recoiled by neutrons and LS detector is set to detect the neutrons scattered from the BGO detector. 
The energy of recoil 
nucleus $E_{\rm R}$ is correlated with the neutron scattering angles by a function given by:
\begin{eqnarray}
E_{\rm R} = E_{\rm n} \left\{ 1 - \left(\frac{m_{\rm n} \cos\theta + \sqrt{m_{\rm N}^2 - m_{\rm n}^2 \sin^2\theta }}{m_{\rm n} + m_{\rm N}}\right)^2 \right\},
\end{eqnarray}
where $E_{\rm n}$ is the energy of the incident neutron, and $m_{\rm n}$ and $m_{\rm N}$ are neutron and nuclei mass, respectively. We have measured the $QF$ for various energies of recoil nuclei. The energies are selected by changing the position of the LS detector at different angles with respect to the BGO detector.

\subsection{Detector configuration}
We have used a BGO scintillation detector composed of a BGO crystal (density of 7.13 g/cm$^3$) with depth and radius of 2 inches each. It is connected to phototubes (R6231-100, Hamamatsu) on both sides and photo-shielded by aluminized Mylar and black sheet. Typically, a light yield of 1,000 photons/MeV is achieved for the energy deposition of charged particles.

To identify the neutron scattering events, we have used a liquid scintillation detector. It consists of a vessel with depth and radius of 3 inches each filled with a liquid scintillator (BC501A, Bicron) and phototube (9822B, ET Enterprises) connected to the window of the vessel.

\subsection{Front-end electronics and data acquisition}

\begin{figure}[htbp]
    \centering 
    \includegraphics[width=0.9\textwidth]
    {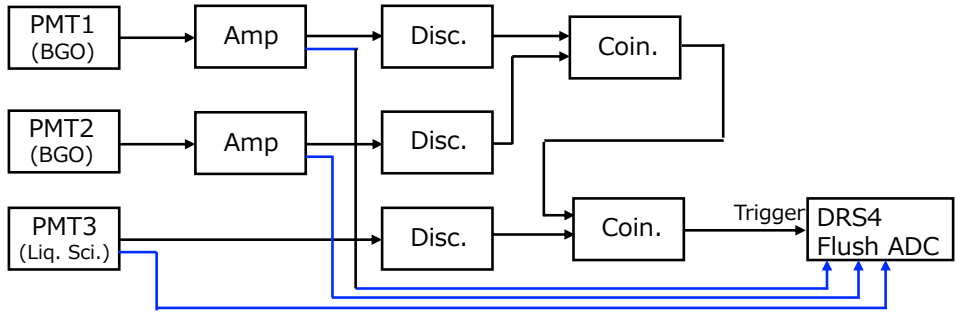}
    \caption{\label{fig:electronics}
    Schematic diagram of front-end electronics and data acquisition.
    }
    \end{figure}

The front-end electronics is shown in Fig.~\ref{fig:electronics}. The waveforms of analog signals are recorded by the flash ADC when both BGO and LS detectors detect signals. The coincidence signals from both phototubes on the BGO detector are required to suppress dark noise and lower the threshold to the level below 1~p.e. for each phototube. Therefore, these thresholds are set to a voltage of 0.5~p.e. equivalent. The width of coincidence signal is set to 200~ns. A trigger signal is generated by the coincidence signals from the BGO and LS detectors. The trigger is fed to the flash ADC (DRS4, PSI) and the waveform of analog signals from these detectors are stored with a precision of 14~bits ADC for a particular pulse height and 
0.7~GSPS\footnote{GSPS: an unit of sampling rate, Gigasamples per second.}
sampling rate for a full range of 1463~ns.

\subsection{Calibration and performance check}
\label{sec:calibration}

\begin{figure}[htbp]
    \centering 
    \includegraphics[width=0.8\textwidth]
    {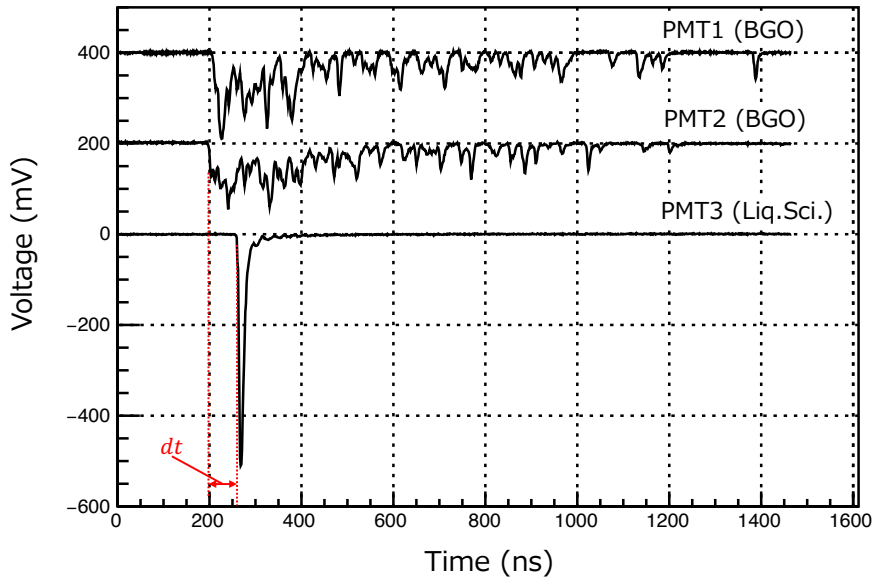}
    \caption{\label{fig:waveform}
    Typical waveforms of the phototubes from the BGO and liquid scintillation detectors.
    }
    \end{figure}

Typical waveforms of the signals are displayed in Fig.~\ref{fig:waveform}. The waveform of PMT 1 and 2 of the BGO detector are shown by shifting +400~mV and +200~mV in this figure, respectively. The expected energy of the recoil nuclei signals is as low as a few dozen to several hundred keV. With such a low energy, the waveform looks like a jagged line made of several single photoelectron signals following the time distribution of an exponential function with a decay constant of 300~ns. The energy deposit in the BGO scintillator is evaluated with the cumulative charge of two PMTs that is an integration of the waveforms from 143 to 1463~ns. 
The waveform of the LS detector is not shifted in Fig.~\ref{fig:waveform}. 
As most of the signal is distributed between 0 and 571~ns 
and the peak width is narrow ($\sim$30~ns), the charge is calculated with an integration window of 571~ns. 
However, the waveform overlapping with the other neutron signals was not observed in this experiment. The arrival time difference $dt$ is calculated as $dt=t_{LS} - t_{BGO}$ where $t_{LS}$ is the rise time of the waveform from the LS detector and $t_{BGO}$ is an average of the rise times for two phototubes of the BGO detector. The rise time is defined as a time when an absolute voltage is over one-third of the pulse height from the baseline.

\begin{figure}[htbp]
    \centering 
    \includegraphics[width=0.8\textwidth]
    {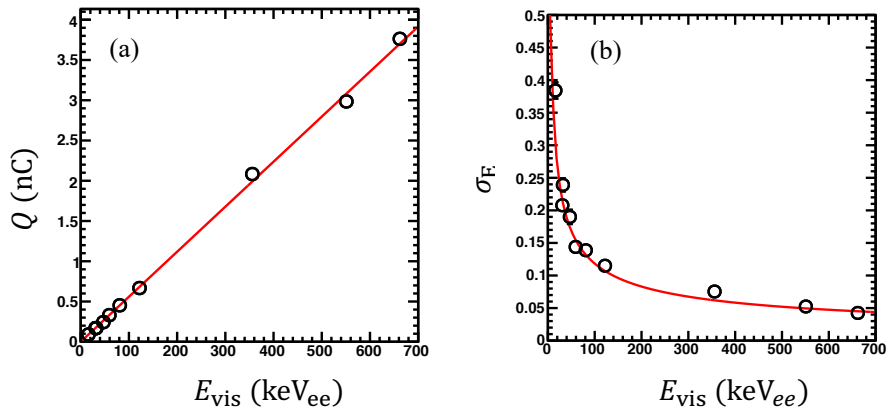}
    \caption{\label{fig:ene.calib}
     Relation between the integrated charge of the signal waveform and energy for BGO detector (a). The evaluated energy resolution as a function of visible energy (b).
     Open circles are data and red lines are a fit function.
     Temperature has been stable at 18 $^\text{o}\rm C$ during the measurement.
    }
    \end{figure}

Energy calibration for the BGO detector was performed using various gamma-ray sources for 
the order of ten to hundreds keV,
including $^{57}$Co (47 and 122~keV), $^{133}$Ba (31, 81, and 356~keV), $^{137}$Cs (32 and 662~keV), and $^{241}$Am (16.1 and 59.5~keV). We have determined the calibration coefficient as $(1.788\pm0.001)\times10^2~{\rm keV_{ee}/nC}$ \footnote{${\rm keV_{ee}}$ is a unit of visible energy assuming energy deposition of electrons in the scintillation detectors.} by a linear fit to the relation between the sum of the integrated charges from two phototubes on both ends of the BGO crystal and gamma-ray energy
as shown in Fig.~\ref{fig:ene.calib}~(a). 
The energy resolution for deposited energy $E_{\rm vis}$ is evaluated as shown in Fig.~\ref{fig:ene.calib}~(b), here, it is defined as the width of one sigma of the Gaussian fit on each photo-electron peak.
The resolution curve is consistent with an inversely proportional function $\sqrt{E_{\rm vis}}$. These indicate the energy threshold of $\sim$7~${\rm keV_{ee}}$ and lower limit of the energy peak at 16~${\rm keV_{ee}}$ with a resolution of 38\%.

\begin{figure}[htbp]
    \centering 
    \includegraphics[width=0.5\textwidth]
    {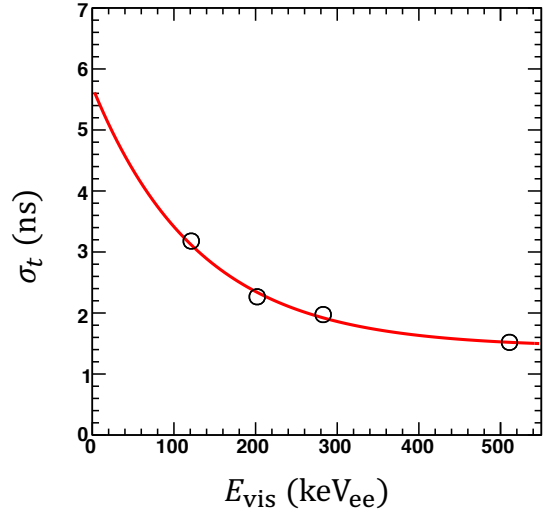}
    \caption{\label{fig:time.reso}
    Timing resolution as a function of visible energy.
    Black open circles are the data.
    Red line is a fit result by a function composed of exponential and constant terms.
    }
\end{figure}

The timing resolution of our detector is evaluated with $^{22}$Na source by applying its positron annihilation gamma rays. The result is shown in Fig.~\ref{fig:time.reso}. It was observed that the timing resolution worsens at lower $\rm keV_{ee}$ as expected, and the resolution is estimated by the exponential fit in the energy region less than hundred $\rm{keV}_{ee}$.

\subsection{Energy and intensity of irradiating neutrons}

\begin{figure}[htbp]
    \centering 
    \includegraphics[width=\textwidth]
    {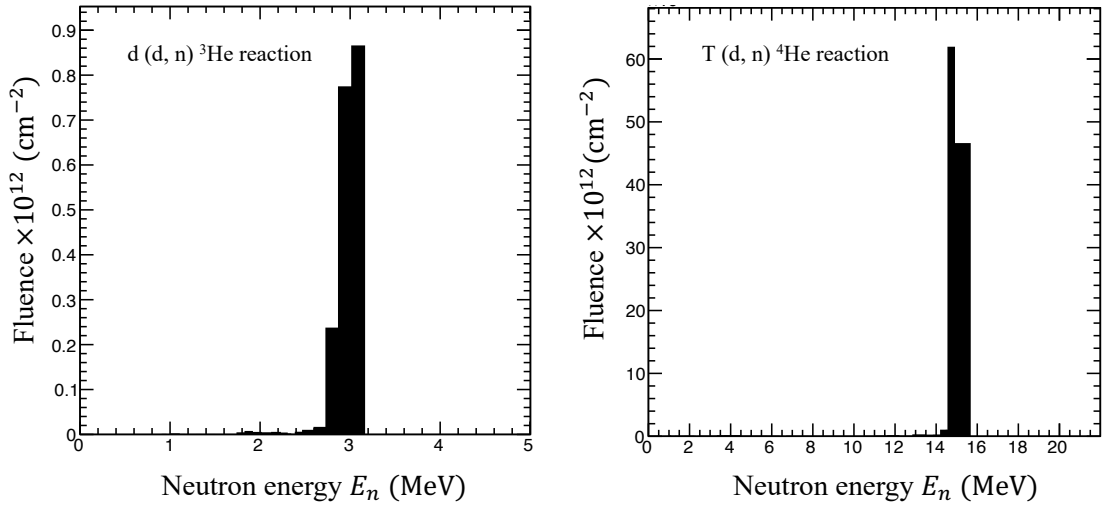}
    \caption{
    Neutron energy spectra
    for $d(d,n)^{3}{\rm He}$ (left) and ${\rm T}(d,n)^{4}{\rm He}$ (right) 
    reactions.
    }
    \label{fig:beam.energy}
    \end{figure}

Energy of the irradiating neutrons
are monitored by the $^3$He detector in a polyethylene Bonner sphere placed at the same position as the BGO detector at a distance of 1 m away from the beam target. Figure~\ref{fig:beam.energy} shows the monitored energy spectra of neutrons generated from ${\rm T}(d,n)^{4}{\rm He}$ and $d(d,n)^{3}{\rm He}$ reactions. The measured spectrum has been adopted in the simulation as the initial neutron energy distributions.

The intensity of irradiating neutrons
is relevant to a ratio of the number of signal events to the accidental coincident background. To evaluate the impact, we have measured two different intensities $\rm{\sim670}$ and $\rm{\sim200}~cm^{-2}~s^{-1}$ for the neutron energy of 14.8~MeV. For 3~MeV neutron irradiation, we have considered the data with the neutron intensity of $\rm{\sim100}~cm^{-2}~s^{-1}$.

\section{Monte Carlo Simulation}
\label{sec:sim}

In this study, we have performed Monte Carlo simulation using GEANT4 toolkit~\cite{GEANT4} (Geant4.9.6p04 + G4NDL4.2) and compared the results with that of the experiment. Geometric configuration of the experimental hall was reproduced in the simulation with a beam pipe, detector stage, wall, mesh floor, and ceiling in AIST. 
The detector, composed of phototubes, BGO crystal, and liquid scintillator, was irradiated by the neutrons.
The detector response was simulated based on the measured energy resolution described in Sec.~\ref{sec:calibration}. The neutron dataset of ENDF-VIII.0 model was adopted for the simulation. Model dependence of the cross sections for elastic and inelastic scattering of oxygen, germanium, and bismuth is negligible while determining $QF$ in this study. The neutron energy distribution shown in Fig.~\ref{fig:beam.energy} was used as the input to the simulation.

In this study, the energy of recoil nuclei is determined by the scattering angle of neutrons. Therefore, bias of the scattering angle $\theta$ due to the secondary scattering in other materials surrounding the detectors causes a shift of the recoil energy, thus resulting in the variation of the measured $QF$. In addition, width of the initial neutron energy causes variation in the energy of recoil nuclei $E_R$ in the detector. These effects were evaluated and accounted for as systematic uncertainty on the dependence of $E_R$ in the $QF$measurement. Details of the systematic uncertainty will be explained in Sec.~\ref{sec.sys}.

\section{Analysis}
\label{sec:ana}

This section presents the procedural analysis on various factors, such as systematic uncertainty, $QF$, energy distribution, and difference between neutron and gamma rays. Elastic scattering events have been selected based on the criteria of neutron identification in the LS detector and relative timing cuts for the BGO and LS detectors. To evaluate net signal events, contamination of the background events was evaluated from the off-time coincidence sample and compared with the events in the signal time window after normalization. Then, the $QF$ was determined by a ratio of the measured peak energy to the estimation of recoil energy.

\subsection{Discrimination of neutron and gamma ray}

\begin{figure}[htbp]
    \centering 
    \includegraphics[width=\textwidth]
    {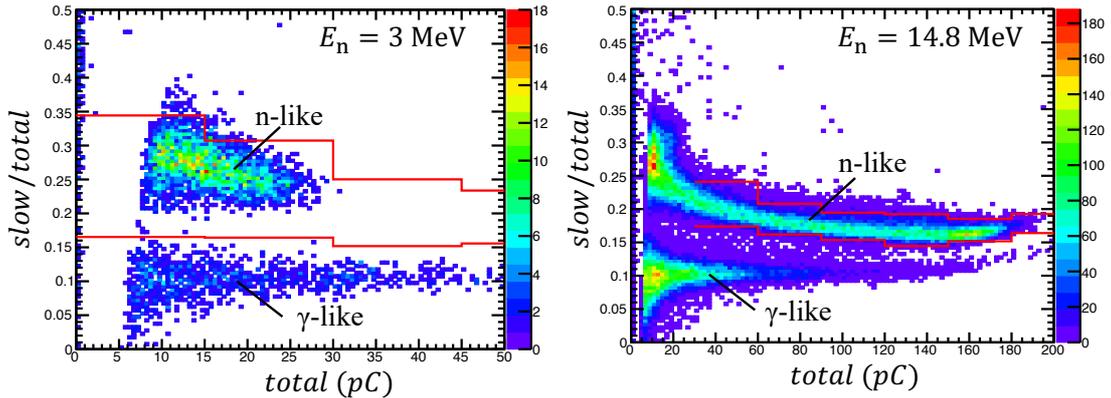}
    \caption{
    Relation between $total$ and $slow/total$ for neutron energies of 3.0 and 14.8~MeV.
    Color at the pixel indicates the count number of events.
    Red lines are the boundaries to discriminate neutron and gamma-ray events.
    }
    \label{figure.PSDs}
    \end{figure}

Neutrons and gamma rays are discriminated by the signal waveforms of the LS detector and are known as pulse shape discrimination (PSD). Figure~\ref{figure.PSDs} shows the relation between ${total}$ and ${slow}/{total}$ for neutron energies of 3 and 14.8~MeV where ${total}$ and ${slow}$ are the integrated charges of the waveform in the nominal time window and shorter time window, respectively. The latter starts 21 ns later. The neutron-like and gamma-like events are distributed in isolation. The boundaries are represented as red lines as shown in Fig.~\ref{figure.PSDs} to discriminate between the two events.

\subsection{Neutron time of flight}

\begin{figure}[htbp]
    \centering 
    \includegraphics[width=0.8\textwidth]
    {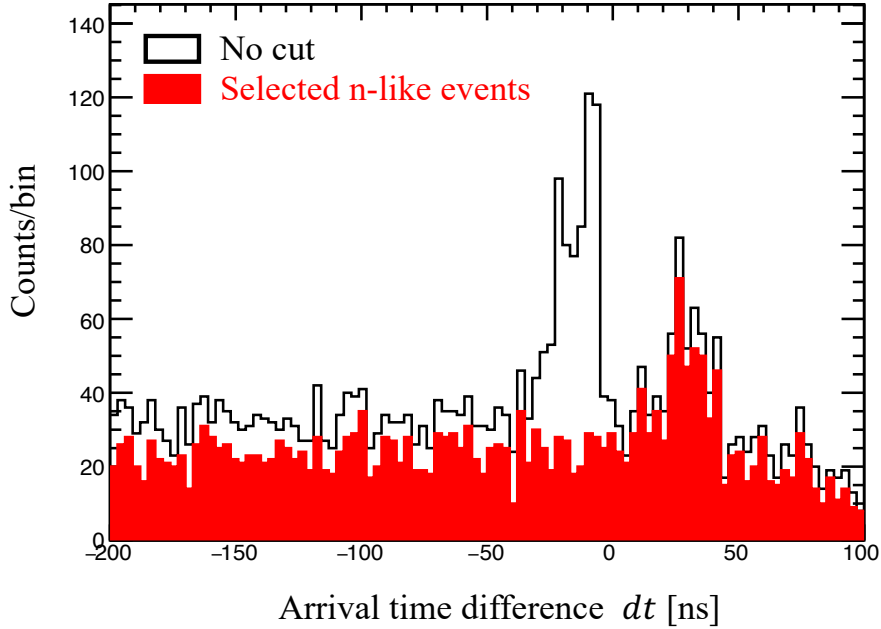}
    \caption{
    Distributions of arrival time difference ($dt$)
    for $E_n=3~\rm MeV$ and $\theta=90^{\circ}$.
    Black open and red filled histograms distributions before and after the PSD cut are applied to select neutron-like events, respectively.
    On-time window is set from $+5$ to $+45$~ns and the off-time window is set from $-200$ to $0$~ns and from $+50$ to $+100$~ns.
    }
    \label{figure.time.diff}
    \end{figure}

Figure~\ref{figure.time.diff} shows the distributions of the arrival time difference ($dt$) when the LS detector is set at $\theta=90^{\circ}$ and the detectors are irradiated to neutrons with $E_n=3~\rm MeV$. The peak value is obtained at approximately 0 ns due to gamma rays induced by inelastic scattering of neutrons in the BGO crystal. After the PSD cut, the remaining peak is identified at approximately 40~ns. Since the scattered neutrons have $\sim$3.0~MeV of kinetic energy and distance between BGO and LS is 1~m, $\sim$50~ns of additional time than gamma rays is required to reach the LS detector after being scattered in the BGO detector. However, a flat component of $dt$ indicates the existence of accidental coincidence of neutrons in the BGO and LS detectors.

\subsection{Energy distribution in BGO}

\begin{figure}[htbp]
    \centering 
    \includegraphics[width=0.8\textwidth]
    {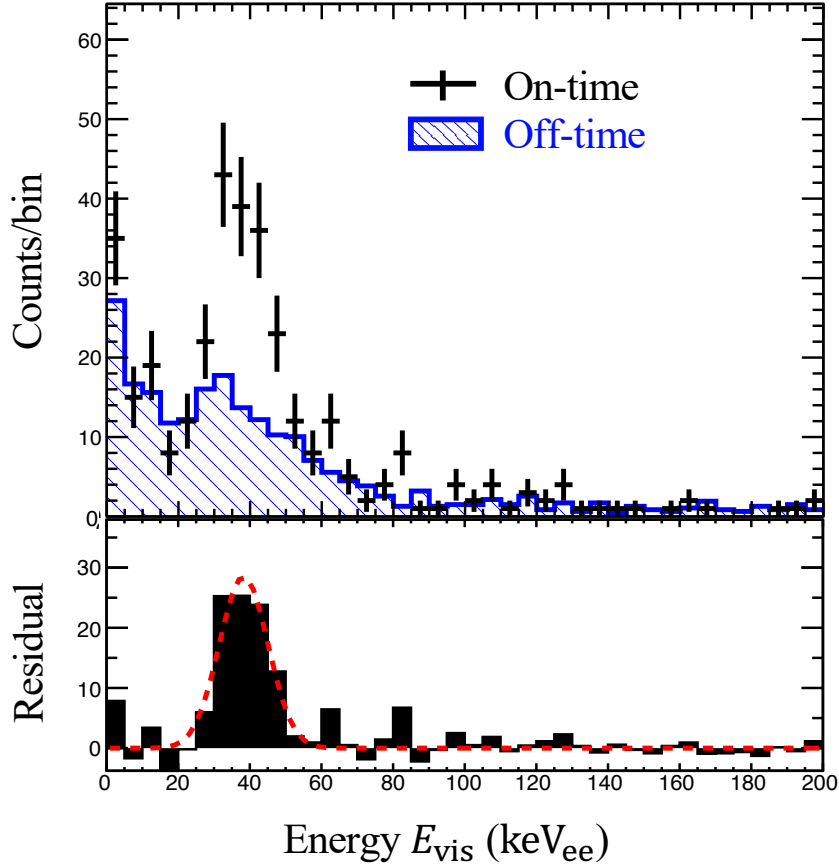}
    \caption{\label{fig:ene.distribution}
    Energy distributions from on-time (black dots) and off-time (blue shade histogram) sample
    for $E_n=3.0~\rm MeV$ and $\theta=90^{\circ}$ (top panel).
    Bottom panel is the residual with
    a Gaussian fit represented by the red dashed line.
    }
    \label{figure.BGO.ene.spectrum}
    \end{figure}

Significant improvement in performance was recorded in the $dt$ distributions of the neutron-like events in various configurations. The signal window was set from $+5$ to $+45$ ns (on-time) to include the elastic scattering events. The background events due to the accidental coincidence were evaluated with two time-windows set at both sides of on-time window; that is, one from $-200$ to 0~ns and the other from $+50$ to $+100$~ns (off-time). These time windows are set for $E_n=3.0~\rm MeV$ while in the case of $E_n=14.8~\rm MeV$, the on-time window was set from -40 to +30 ns, and the off-time window was set at both sides as $-200$ to $-40$~ns and $+30$ to $+100$~ns, respectively. The number of events in the off-time sample was normalized by the ratio of the ranges of on-time to off-time windows for a comparison with the on-time sample. Figure~\ref{figure.BGO.ene.spectrum} shows the energy distributions from on-time and off-time samples observed in the BGO detector. We found a significantly excess value with a peak at approximately $E_{\rm vis}=40~\rm keV_{ee}$, which is due to oxygen recoil. For $E_n=3.0~\rm MeV$ and $\theta=90^{\circ}$, the recoil energy of oxygen nuclei was calculated as $E_R = 334 \pm 17~\rm keV$ where the uncertainty includes the width of the neutron energy and range of the scattering angle due to the finite size of the BGO crystal and LS container.

\subsection{Quenching factor determination}

$QF$ is defined as $QF=E_{\rm vis} / E_R$. The visible energy for the recoil oxygen is given as $E_{\rm vis}=38.5\pm1.3~\rm keV_{ee}$ from the Gaussian fit to the peak. Then, $QF$ was evaluated to be $(11.5\pm0.4)\times10^{-2}$ from the value of $E_R = 334~\rm keV$ for a configuration of $E_n=3.0~\rm MeV$ and $\theta=90^{\circ}$.

Further, we have performed another approach to determine the $QF$ more precisely. The measured energy spectra in the on-time sample were fitted by a combination of the simulated energy distribution for oxygen recoil and background distribution in the off-time sample with three free parameters: 1) normalization factor for the simulated energy distribution for oxygen recoil; 2) $QF$ value to change the energy scale of simulation; and 3) normalization factor for the off-time distribution. The minimum $\chi^2$ value, $\chi^2/dof = 25.4/37$, was obtained with $QF = (11.49^{+0.59}_{-0.41})\times10^{-2}$ in the fit for $E_R = 334~\rm keV$. The value of minimum $\chi^2$ is consistent with the best fit value of $QF$.

\begin{figure}[htbp]
    \centering 
    \includegraphics[width=\textwidth]
    {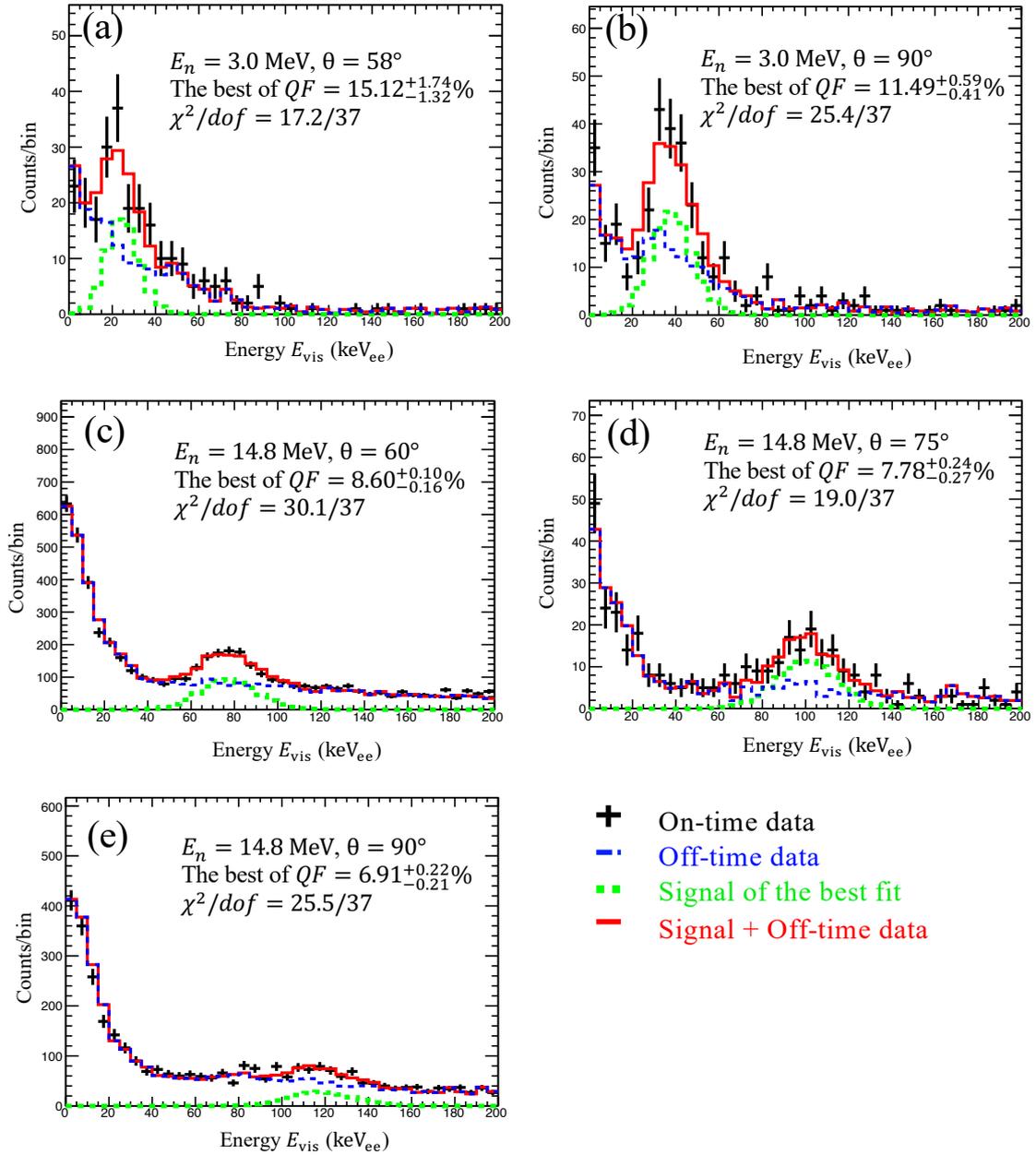}
    \caption{\label{fig:QF}
    Fit results for five different configurations.
    Neutron energy $E_n$ and angle of the LS detector $\theta$ are specified in each plot.
    Black dots are the on-time data overlaid with the off-time data (blue dashed-line), simulated energy distribution for elastic scattering scaled by the best fit value of $QF$ (green dotted line), and combination of off-time data and simulated spectrum (red solid lines). Simulated spectra and off-time data are normalized with the best fit parameters. 
    }
    \label{figure.fit.result}
    \end{figure}

Figure~\ref{figure.fit.result} shows the fit results for five configurations that generate different energies for recoil oxygen:
a) $E_n=3.0~\rm MeV$, $\theta=58^{\circ}$, and $E_R=163~\rm keV$,
b) $E_n=3.0~\rm MeV$, $\theta=90^{\circ}$, and $E_R=334~\rm keV$, 
c) $E_n=14.8~\rm MeV$, $\theta=60^{\circ}$, and $E_R=899~\rm keV$, 
d) $E_n=14.8~\rm MeV$, $\theta=75^{\circ}$, and $E_R=1305~\rm keV$,
e) $E_n=14.8~\rm MeV$, $\theta=90^{\circ}$, and $E_R=1726~\rm keV$. 
Quenching factors are determined by analyzing the fit to each distribution. The observed spectra in the on-time sample are in good agreement with the combination of off-time data and simulated spectrum for the recoil oxygen nuclei if the quenching factors are applied.

\subsection{Systematic uncertainties}
\label{sec.sys}

In this study, we have considered systematic uncertainties to determine the $QF$ of oxygen recoil energy. The main sources of uncertainties are off-time window selection and energy linearity. For the former source, the variation of $QF$ was evaluated with the off-time window expanded by a factor of $\sim2$ and was computed as 0.87\% in the setup of $\theta=90^{\circ}$ in $E_n=3.0~\rm MeV$. For the latter source, the uncertainty was estimated from the fit error in the calibration (see Fig.~\ref{fig:ene.calib}) as 6.8\% in the same setup. Uncertainties for the other setups of $\theta$ and $E_n$ were estimated using this approach. It was observed that the total systematic uncertainties in the $QF$ measurement were comparable to the statistical uncertainties for all setups. In addition, the uncertainties in the recoil energy ($E_R$) were also accounted. As the main sources, uncertainties due to the detector volume and initial neutron energy width were estimated with simulation. For the setup of $\theta=90^{\circ}$ in $E_n=3.0~\rm MeV$, total uncertainty on $E_R$ was evaluated as $\pm17$~keV for 334~keV. Systematic uncertainties for all setups are summarized in Table~\ref{tab:QF.result}.

\section{Results}
\label{sec:result}

\begin{figure}[htbp]
    \centering 
    \includegraphics[width=0.8\textwidth]
    {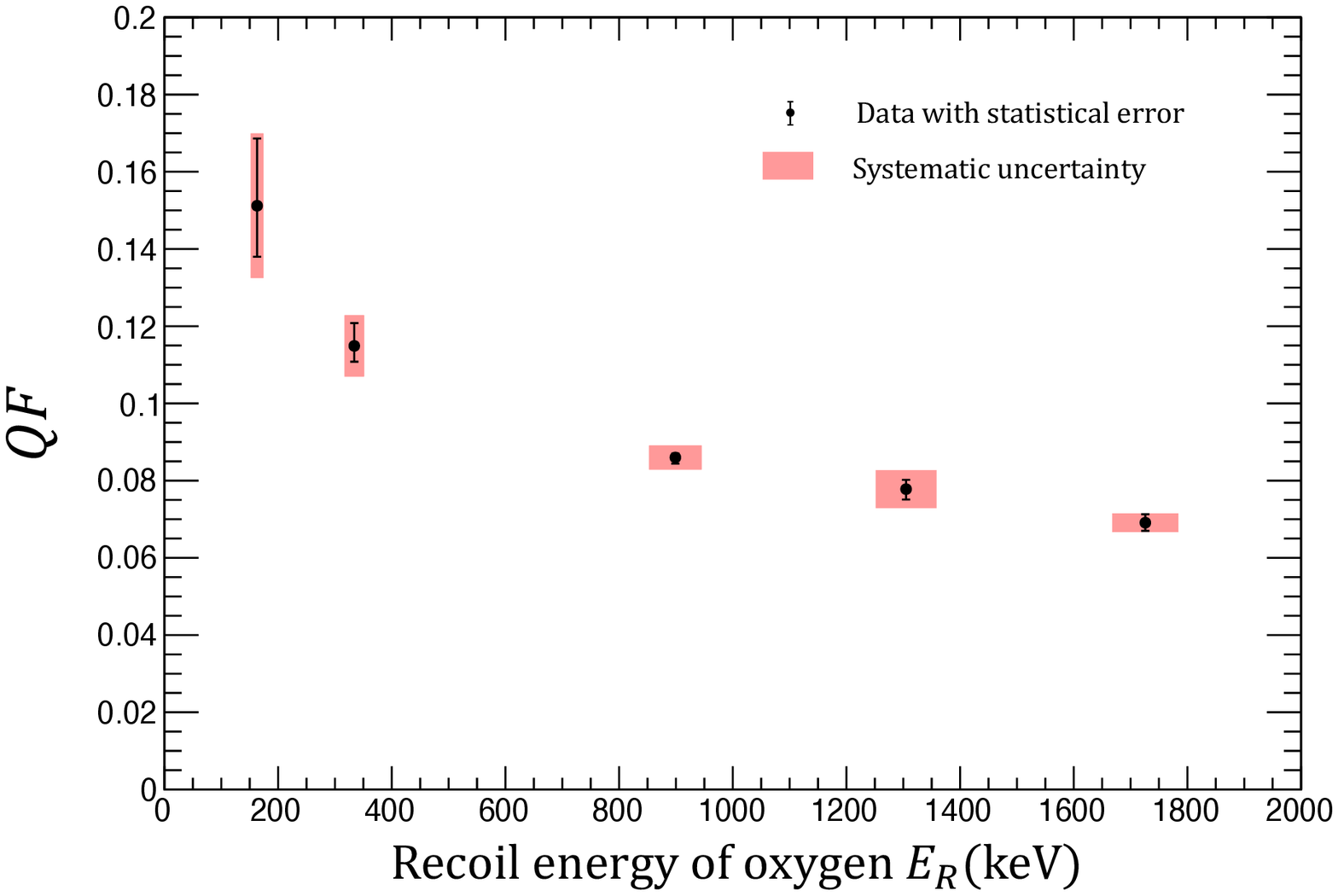}
    \caption{\label{fig:QF}
    Quenching factor as a function of oxygen recoil energy.
    Black dots are the data with the statistical uncertainties (black bar) and systematic uncertainties (red band).
    }
    \label{figure.QF.as.function}
    \end{figure}

The measured $QF$ values and uncertainties are summarized in Table~\ref{tab:QF.result} and also shown in Fig.~\ref{figure.QF.as.function} as a function of nuclear recoil energy ($E_R$). The energy dependence of $QF$, especially the increase at lower energy, was clearly observed.

\begin{table}[htbp]
\center
\caption{
Summary of quenching factor measurements for oxygen recoil in BGO crystals.
The scattering angle $\theta$, initial neutron energy $E_{\rm n}$, and recoil energy $E_{\rm R}$ are listed with the measured values of the visible energy $E_{\rm vis}$ and quenching factor $QF$ for oxygen nuclei.
}
\begin{tabular}{ccccc}
\hline
 $E_{\rm n}$~(MeV) &$\theta$~(deg.) &$E_{\rm R}$~(keV) 
 & $E_{\rm vis}~(\rm keV_{ee})$
 & $QF\times10^2$\\
\hline
 3.0  & $58\pm1$ & $163\pm11$ & 25.2$\pm$2.4 & $15.12^{+1.74}_{-1.32}(\rm stat.)\pm1.87(\rm syst.)$ \\
      & $90\pm1$ & $334\pm17$ & 38.6$\pm$5.8 & $11.49^{+0.59}_{-0.41}(\rm stat.)\pm0.79(\rm syst.)$ \\
 14.8 & $60\pm1$ & $899\pm46$ & 76.9$\pm$1.1 & $8.60^{+0.10}_{-0.16}(\rm stat.)\pm0.31(\rm syst.)$ \\
      & $75\pm1$ & $1305\pm53$ & 96.6$\pm$2.1 & $7.78^{+0.24}_{-0.27}(\rm stat.)\pm0.49(\rm syst.)$\\
      & $90\pm1$ & $1726\pm58$ & 116.9$\pm$10.2 & $6.91^{+0.22}_{-0.21}(\rm stat.)\pm0.24(\rm syst.)$\\
\hline
\end{tabular}
\label{tab:QF.result}
\end{table}

\section{Conclusion}
\label{sec:Conclusion}

In this study, we conducted an experiment to measure the $QF$ of recoil energy for oxygen nuclei in BGO scintillators by the irradiation of monochromatic neutrons. The $QF$ measured at different nuclei recoil energies was in the range of 163–1726~keV and the significant energy dependence was confirmed.

\acknowledgments
This work is supported by 
the Neutron measurement consortium for Underground Physics and
JSPS KAKENHI Grants Grant-in-Aid for Scientific Research (C) No. 20K03998 and Grant-in-Aid for Scientific Research on Innovative Areas 19H05808.
We would like to thanks to National Institute of Advanced Industrial Science and Technology (AIST), Japan for providing the well-controlled neutron irradiation.

\bibliography{Reference.bib}

\end{document}